\documentclass[useAMS,usenatbib]{mn2e}
\usepackage{graphicx}
\usepackage[latin1]{inputenc}
\usepackage{subfigure}
\usepackage{indentfirst}
\usepackage{multicol}
\usepackage{Times} 
\usepackage{url}

\title[Stable orbits around the triple system Eugenia]
{Mapping stable direct and retrograde orbits around the triple system of asteroids (45) Eugenia}
\author[Araujo, R.A.N.; Moraes, R.V; Prado, A.F.B.A; Winter, O.C.]
{R.A.N.Araujo$^{1,2,3}$\thanks{E-mail:ran.araujo@gmail.com};  Moraes, R.V.$^{2}$; A.F.B.A.Prado$^{1}$; O.C.Winter$^{3}$  \\
$^1$INPE - National Institute for Space Research, CEP 12201-970, S\~ao Jos\'e dos Campos, SP, Brazil\\
$^2$  UNIFESP/ICT-Federal University of S\~ao Paulo, CEP 12 247-014, S\~ao Jos\'e dos Campos, SP, Brazil \\
$^3$UNESP - Univ Estadual Paulista, Grupo de Din\^amica Orbital e Planetologia, CEP 12516-410, Guaratinguet\'a, SP, Brazil.}
\begin{document}

\date{}

\pagerange{\pageref{firstpage}--\pageref{lastpage}} \pubyear{2017}
\maketitle
\label{firstpage}

\begin{abstract}
It is well accepted that knowing the composition and the orbital evolution of asteroids may help us to
understand the process of formation of the Solar System.
It is also known that asteroids can represent a threat to our planet.
Such important role made space missions to asteroids a very popular topic in the current astrodynamics and astronomy studies.
By taking into account the increasingly interest in space missions to asteroids, especially to multiple systems, we present a study aimed to 
characterize the stable and unstable regions around the triple system of asteroids (45) Eugenia. 
The goal is to characterize unstable and stable regions of this system and compare with 
the system 2001 SN263 - the target of the ASTER mission.
Besides, \cite{prado2014} used a new concept for mapping orbits considering the disturbance 
received by the spacecraft from all the perturbing forces individually.
This method was also applied to (45) Eugenia.
We present the stable and unstable regions for particles with relative inclination between $0^{\circ}$ and $180^{\circ}$. We found that (45) Eugenia 
presents larger stable regions for both, prograde and retrograde cases. This is mainly because the satellites of this system are small when compared 
to the primary body, and because they are not so close to each other.
We also present a comparison between those two triple systems, and a discussion on how these results may guide us in the planning of future missions.

\end{abstract}


\section{Introduction}

\begin{table*}
\label{tab_elements}
\begin{minipage}{115mm}
\centering
\caption{Physical data and orbital elements of the  components of the triple system (45) Eugenia $^{(1)}$}
\end{minipage}
\begin{tabular}{ccccccc}
\hline 
Body		&Orbits		&$a$		&$e$		&$i$ $^{(2)}$		&Radius (km)		&Mass (kg)  		       \\ 
\hline	
Eugenia		&Sun		&$2.72$ au	&$0.083$	&$18.2^{\circ}$		&$108.5$$^{(3)}$  	&$5.63\times10^{18}$  	       \\\\
Princesse	&Eugenia	&$610.8$ km	&$0.069$	&$18^{\circ}$		&$2.5$  		&$2.5\times10^{14}$   	       \\\\
Petit-Prince	&Eugenia	&$1164.5$ km	&$0.006$	&$9^{\circ}$		&$3.5$  		&$2.5\times10^{14}$  	       \\\\
\hline
\multicolumn{7}{l}{(1) \cite{marchis} and JPL's Horizons system, for the epoch Julian date 2452980.0 JD.}\\
\multicolumn{7}{l}{(2) Inclinations of Petit-Prince and Princesse with respect to the equator of Eugenia and }\\
\multicolumn{7}{l}{inclination of Eugenia relative to the heliocentric ecliptic.}\\
\multicolumn{7}{l}{(3) Equivalent radius \citep{marchis}.}\\
\end{tabular}
\end{table*}

To know the composition and the dynamics of the asteroids may help us 
to understand the process of formation of the Solar System and, so, the formation and composition of our own planet. 
Besides, it is also known that asteroids can represent a threat to our planet.
These are some of the reasons why space missions aimed to visit asteroids in the Solar System is a very popular topic in current astrodynamics and astronomy studies. 

Examples of some successful missions are the OSIRIS-REx (NASA), launched in September, 2016, and that must return to the Earth with a sample of the 
NEA (101955) Bennu \citep{b15}, or the Hayabusa mission (JAXA), launched in May, 2003, which explored the NEA (25413) Itokawa \citep{b10} and returned to 
Earth in June, 2010, with samples from the surface of the asteroid \citep{fujita}.

Multiple systems of asteroids are interesting targets for space missions, since they increase the range of possible scientific investigations. 
The ASTER mission  - the First Brazilian Deep Space Mission \citep{sukhanov}, was concepted taking this advantage into account.
In that sense, searching for stable orbits around those bodies is very important, both in terms of science and engineering. 
Regions of stable orbits may be places full of dust and/or small pieces of materials, as well as possible locations for other members of the system. 
In a mission, they may also indicate good regions to place a spacecraft to observe the system. 

The primary target of the ASTER mission is the NEA 2001 SN263, which is a triple system of asteroids. 
The announcement of this mission has motivated studies aimed to characterize regions of stability of this system. 
\cite{araujo2012} and \cite{araujo2015}, characterized stable and unstable regions around the components of this triple system, 
for the prograde and retrograde cases, through numerical integrations of the gravitational N-body problem. 
\cite{prado2014} mapped orbits for a small body orbiting the asteroid 2001 SN263. He considered the disturbance 
received from all the perturbing forces individually. This study used a new concept for mapping orbits that 
shows the relative importance of each force for a given orbit in the system. Such information helped to make a decision about which forces 
need to be included in the model for a given accuracy and nominal orbit. 

Thus, the present paper uses a combination of both methods presented in  \cite{araujo2012}, \cite{araujo2015} and \cite{prado2014}, 
in order to search for direct and retrograde stable orbits in the (45) Eugenia triple system of asteroids. 
(45) Eugenia is a triple system of asteroids of the MBA (Main Belt Asteroids). 
It is composed by the central body Eugenia and by the satellites Petit-Prince and Princesse. Physical characteristics and orbital elements of this system 
are presented at Tab. $\ref{tab_elements}$. The physical and orbital data for Eugenia and its satellites were obtained from 
\cite{marchis}. \cite{Beauvalet2014} updadet those data with minor corrections in the order of the error bar of the first work. 
Although those corrections must be important from the observational point of view, they do not significantly affect our long-term analysis of the 
stability of the particles within the system.

In particular, in this paper we searched for regions where direct orbits are unstable, but retrograde orbits are stable. 
Those orbits are very good for a potential mission.
The probe can benefit of the stability of the orbit to minimize station-keeping maneuvers and, at the same time, 
to travel in regions that are expected to be free of dust, so reducing the risk of collisions with natural debris. 
This idea was proposed before in the literature for the system 2001 SN263 \cite{araujo2015}, and is now extended to the (45) Eugenia system. 

The main reason for this extension is that those two systems are completely different from each other in terms of physical characteristics. 
The 2001 SN263 is much smaller in sizes and distances, with the bodies very close to each other. 
It means that the possible orbits for the spatial exploration of the system are strongly perturbed and stable orbits are very rare in many important regions. 
The (45) Eugenia has components located at much larger distances, 
so the general mappings of the stable orbits are very much different. Even the Kozai effects are reduced, allowing inclined orbits 
in many situations where they did not appear in the system 2001 SN263. It is easy to quantity the differences, in terms 
of the perturbation received by a small body in both systems, using the integral of the perturbing forces for one orbital period \citep{prado2013}.

The approach used here to find stable orbits is similar to the one used in the previous 
studies related to the 2001 SN263 system, and it is described in Sec. \ref{sec_method}

\section{Method}
\label{sec_method}

\begin{figure}
\begin{center}
\includegraphics[scale=0.32]{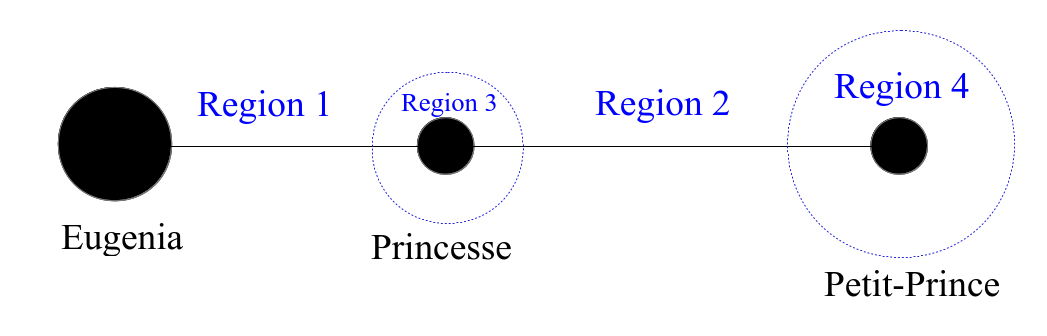}
\caption{Representation of the triple system 45 Eugenia and the regions of influence of each component.}
\label{fig_encounter}
\end{center}
\end{figure}

The method adopted consisted in dividing the regions around the triple system 
in four distinct internal regions: two regions around the two small satellites of the system and the regions between the orbits 
of the two satellites (see Fig. \ref{fig_encounter}). 

Regions $3$ and $4$ around the satellites Princesse and Petit-Prince were defined by the Hill's radius \citep{murray}, 
considering an approximation given by the two-body problem between Eugenia and each one of the satellites 
separately. We found a region of about $29$ km where the gravitational perturbation from Petit-Prince is dominant over the 
other bodies of the system. For Princesse this region has a radius of about $15$ km. These values have guided us to define how the particles 
would be distributed within the system. This is better described in subsections \ref{sec_r1} - \ref{sec_r4}

The N-body gravitational problem was integrated in time using the Gauss-Radau numerical integrator \citep{everhart} for a time-span of 2 years. 
Numerical integrations were performed considering a system composed by seven bodies: the Sun, the planets Earth, Mars and Jupiter,
the three components of the asteroid system and thousand of particles randomly distributed around these components, 
including planar and inclined prograde and retrograde orbits. 

Regarding the shape of the body, following \cite{araujo2012} and \cite{araujo2015}, we considered as a first approximation the oblateness of 
Eugenia through the $J_2$ value.
According to \cite{marchis} the $J_2$ value for Eugenia is equal to $0.060 \pm 0.002$ . \cite{Beauvalet2014} fit a dynamical model to 
simulate the observational data and to determine constraints in the dynamical parameter. From this fit they found a 
$J_2$ value equal to $0.0589 \pm 0.0004$. Based on those values and on their error bar, we considered in the integrations $J_2=0.06$.
The difference is less than $2\%$, which do not significantly affect our long-term analysis of stability of particles within the system.

The results are expressed by plots showing the percentage of 
particles that survived, for each set of initial conditions, as a function of the semi-major axis, eccentricity and inclination 
of the initial orbit. The regions where all the particles survived along the time span of 2 years were called stable regions. On the other 
hand, regions where no particles survived along this time span were called unstable regions.

In section \ref{sec_results} we present the results of the numerical integrations for each region and also their implication 
in a space mission, comparing with the results previously found for the triple system 2001 SN263 in the context 
of the ASTER mission.

\section{Results}
\label{sec_results}

\begin{figure}
\begin{center}
\subfigure[]{\includegraphics[scale=0.85]{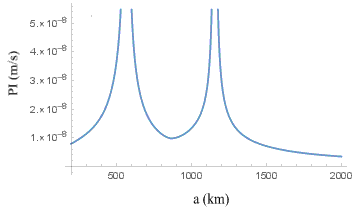}}
\subfigure[]{\includegraphics[scale=0.85]{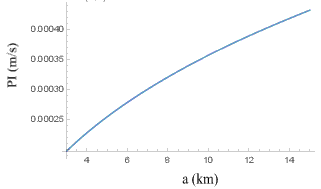}}
\subfigure[]{\includegraphics[scale=0.85]{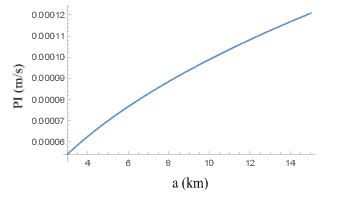}}
\caption{Perturbation Integral $(m/s)$ as a function of the semi-major axis of the orbit of a small body within the triple system, 
considering perturbations coming from a) Princesse and Petit Prince for circular orbits, b) Eugenia and Petit-Prince and c)
Eugenia and Princesse.}
\label{fig_bert}
\end{center}
\end{figure}

A first look at both systems of asteroids indicates that they are quite different in physical terms. 
The system (45) Eugenia has a central body that is much more massive with respect to its satellite bodies, 
when compared with the 2001 SN263 system. In this system the three bodies have a more uniform distribution of masses among the bodies. 
The distances are also different, with the bodies of the system 2001 SN263 being much closer to each other than the bodies that composes 
the (45) Eugenia system. This combination of factors affect 
very much the stability of the orbits, with important consequences in selecting possible orbits for a potential mission having this 
triple system as a target. 
Those situations will be explored in detail when showing the results obtained here.

A previous general study of those differences can be done using the integral of the perturbing forces acting in this system
(\cite{prado2013}, \cite{carvalho2014}, \cite{lara2016}, \cite{olivprado2014}, \cite{oliveiraetal2014}, \cite{sanchesetal2016}, \cite{sanchesetal2014}, 
\cite{santosetal2015}, \cite{shortetal2016}).

Basically, an integral of the perturbing forces acting in a small body within a given system is made over one orbital period. 
The idea is to give a first approximation of the differences of the perturbation level in both systems. 
\cite{prado2014} made this study for the 2001 SN263 system. This paper also showed that the perturbation coming from the two smaller members of 
the system is several orders of magnitude larger than the other perturbations, so it is valid to make a first analysis of the perturbations 
using only those forces. \cite{prado2014} describes well the level of perturbation, showing two peaks near the orbits of the smaller bodies. 
This result is expected, since the perturbation increases when the small body gets closer to the perturbing bodies, 
but this figure quantify the level of perturbation. It is noted that the magnitude of the perturbation is of the order of 0.025 m/s in the regions 
between the orbits of the two smaller bodies. 

The present paper makes a similar analysis for the (45) Eugenia, which results are  shown in Figs \ref{fig_bert}. Since the idea is to measure the effects 
of the perturbations with respect to the gravity field of the main body, which is the most important comparison, a new type of integral is used. 
This index is calculated by the integration of the perturbing forces acting in a specific orbit divided by the gravity field of the main body, over 
one orbital period of the given orbit. The index used in \citep{prado2014} was calculated to compare different forces in the same system, and for orbits 
around the main body, so the effects of dividing the perturbation by the gravity field of the main body are small. But, in the present case, where 
the idea is to compare two different systems, as well as orbits around the main body and the two smaller bodies, it is very important to quantify the level 
of perturbations compared to the gravity field of the main body. 

Figure \ref{fig_bert}a considers orbits around the main body of the system, Eugenia, perturbed by the two natural satellites of the system, Princesse and 
Petit Prince. It is noted the same two 
peaks near the orbits of the perturbing bodies showed in 2001 SN263 system, as expected, but the magnitude of the perturbation levels in all the 
regions are smaller. It is of the order of $10^{-8}~m/s$ in the regions between both smaller bodies, and even smaller after the orbit of the exterior natural satellite. 
The peaks are not shown to keep a better scale, to see more detail, but they are of the order of $10^{-6}~m/s$. The same type of integral index was calculated for the 2001 SN263, 
since \citep{prado2014} uses a slight different index, showing values of the order of $10^{-6}~m/s$ for
the minimum points and near $10^{-5}~m/s$ for the peaks. It clearly indicates the weaker effects of the perturbations, in the (45) Eugenia system, 
from the two companion bodies, for orbits around the main body of the system. This is a mathematical indicative of much less perturbed orbits, 
which will have several consequences on the choice of the locations to place a spacecraft to observe the system. It also shows the importance of using a 
scalar index to quantify the level of perturbation received in a given orbit, so it is not necessary to limit the study to the expected general 
behavior of the perturbation level, without a quantitative analysis of the forces.

The integral of the perturbing forces acting in the system give us an estimation on the stability of the system. It quantify the perturbations due to the 
bodies of the system, indicating regions where particles or a small body  would be more or less perturbed. Nevertheless, it did not take into
account other dynamical effects such as mean motion resonances, the Kozai resonances for high-inclined orbits or the known increase of 
stability of retrograde orbits. The results presented in Fig. $2$ show, as expected, that the perturbation from the satellites is small in the triple
system Eugenia. Thus, we expect the regions around the components of this system to be stable. In order to verify if this system is subject to the 
perturbations cited above and how they differ from what is predicted by the integral method, we performed numerical simulations of particles distributed
within the system and verified their stability as a function of their initial orbit, as follows.

\subsection{Region 1}
\label{sec_r1}

\begin{figure*}
\mbox{%
\subfigure[]{\includegraphics[height=2.9cm]{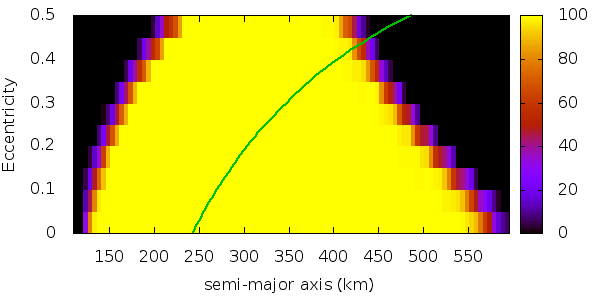}}\qquad
\subfigure[]{\includegraphics[height=2.9cm]{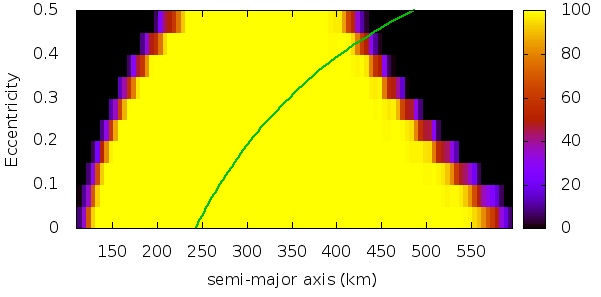}}
\subfigure[]{\includegraphics[height=2.9cm]{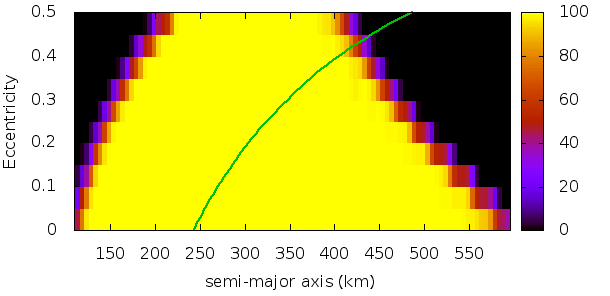}}}

\mbox{%
\subfigure[]{\includegraphics[height=2.9cm]{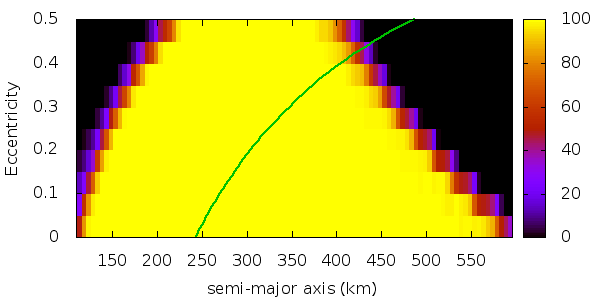}}\qquad
\subfigure[]{\includegraphics[height=2.9cm]{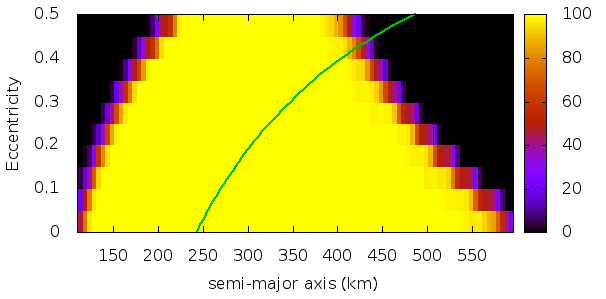}}
\subfigure[]{\includegraphics[height=2.9cm]{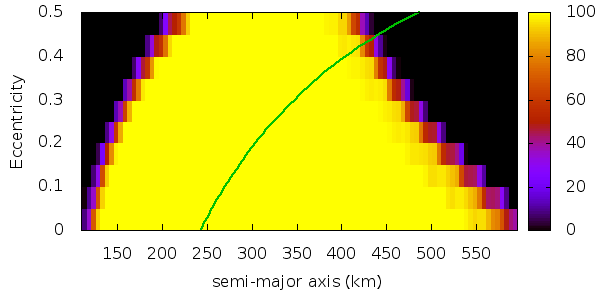}}}

\mbox{%
\subfigure[]{\includegraphics[height=2.8cm]{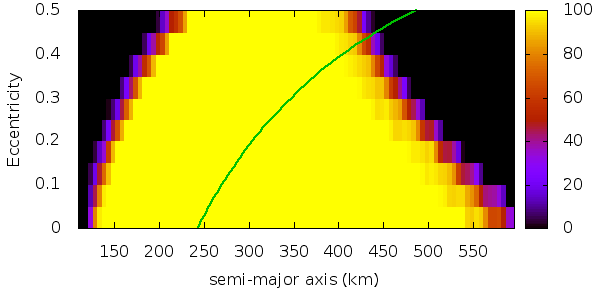}}}
\caption{Diagrams $(a \times e)$ of stability of region $1$ for the triple system (45) Eugenia, for a time span of $2$ years. 
In this region the particles orbit Eugenia with orbital inclination: a) $I=0^{\circ}$, b) $I=30^{\circ}$, c) $I=60^{\circ}$, 
d) $I=90^{\circ}$, e) $I=120^{\circ}$, f) $I=150^{\circ}$, g) $I=180^{\circ}$.
The color-coded scale indicate the percentage of particles that survives in this region, going from $0\%$ (instability) until $100\%$ (stability).
As explained in the text, for a particle with pericentric distance smaller than $243~km$ (on the left of the green lines) a model considering a more 
realistic shape of Eugenia must be adopted.}
\label{fig_histogram1}
\end{figure*}

Region $1$ was defined as the region around Eugenia going from $110$ km ($1.01$ radius of Eugenia) to $595$ km 
(limited by the orbit of Princesse, considering its region of influence). 

The particles were distributed within this region orbiting Eugenia with the following orbital elements:
$110~km~\leq~a~\leq~595~km$, taken every $5~km$, $0\leq e \leq 0.5$, taken every $0.05$, $0^{\circ}~\leq~I~\leq~180^{\circ}$, taken every $30^{\circ}$.
For each combination $(a \times e)$, we considered 100 particles with random values of $f$, $\omega$ and $\Omega$, where $a$ is the semi-major axis,
$e$ is the eccentricity, $I$ is the inclination, $f$ is the true-anomaly, $\omega$ is the argument of the pericenter and $\Omega$ is the longitude of the ascending node. 
This combination of initial conditions resulted in a total of approximately $108,000$ particles distributed within region $1$.

The stability and instability are defined by the number of particles that survived (no ejections or collisions) throughout the numerical integrations. 
The collision was defined by the physical radius of the bodies. For region $1$, an ejection was considered every time that the relative distance 
between a given particle and Eugenia was greater than $595$ km (the limit of the region $1$).

The results for region $1$ is presented in the diagrams of Fig. \ref{fig_histogram1}. 
They show that approximately the whole region is stable. The instability appears only for the orbits really close 
to Eugenia or Princesse, or for those orbits with high eccentricities such that the crossing of orbits become possible, or for those orbits such 
that the pericenter of the orbit is inside Eugenia, leading to collisions.

They also show that there are no effects from the inclination of the orbit in the survival of the particles. 
It is opposite to what was found in the system 2001 SN263, where the inclination played an important role in the evolution of 
the orbits of the particles. \cite{araujo2012} showed, in Fig. $6$, that the Kozai effect reduces the regions of stable orbits 
with the increase of the inclination, with a maximum effect at 60 degrees of inclination. 
Studying retrograde orbits, \cite{araujo2015} showed, in Fig. $8$, that there are locations in the region $1$ with 
stable retrograde and unstable direct orbits. Those regions are excellent locations to place a space probe willing 
to observe the central body, as explained before. 

Therefore, the (45) Eugenia system does not have this option for 
selecting orbits to observe the main body of the system. It happens because the third-body perturbation 
acting in a small body within this system is much smaller in the (45) Eugenia system, when compared to the 2001 SN263 system. This is due to two factors. 
The third-body perturbation depends on the mass ratio between the perturbing and the central body. This ratio is 
(mass of Princesse, the closest perturbing body)/(mass of Eugenia) = $(2.51 \times 10^{14})/(5.63 \times 10^{18}) = 4.46 \times 10^{-5}$ in the case of the (45) 
Eugenia system and $1.06 \times 10^{-2}$ for the 2001 SN263 system, a very large difference. Another point is that both companion asteroids are located 
very far from the main body, more than 600 km for the closest one, as shown in Tab $1$. This distance is about $3.8$ km in the 2001 SN263 system. 
It means that a combination of those two facts make the Kozai effects to be much smaller and not able to 
modify significantly the evolution of planar and inclined orbits.

The particles in region $1$ are those subjected to suffer the effects due to the irregular shape of Eugenia. 
\cite{jiang} explored these effects on the two satellites Princesse and Petit-Prince. 
Nevertheless, \cite{chanut} has shown how the relative error on the gravitational potential depends on the radial distance of a particle relative to 
a central irregular shaped body. In their work the gravitational potential was determined for the asteroid 433 Eros modeled using the polyhedral model.
They compared the results obtained with a simpler model where Eros were considered a point of mass. 
Their results show an error smaller than $10\%$ at a distance of about $1.6$ of the physical semimajor axis of the body. 
Applying this result to Eugenia, considering the physical semimajor axis of Eugenia as $152~km$ \citep{jiang},  we estimate that
particles within the distance of about $243~km$ are those that may suffer the effects of the irregular shape of Eugenia. 
Thus, for a particle with pericentric distance smaller than $243~km$ (on the left of the green lines shown in the plot available in Fig. \ref{fig_histogram1}) 
a model considering a more realistic shape of Eugenia must be adopted. For the other regions the differences are not significant.

\subsection{Region 2}
\label{sec_r2}

Region $2$ is the region between the orbits of the two satellites, and its was filled with particles orbiting Eugenia.
This region goes from $625.8$ km (limited by the orbit of Princesse, considering 
its region of influence) to $1136.5$ km (limited by the orbit of Petit-Prince, considering 
its region of influence). 
The particles were distributed within this region orbiting Eugenia with 
$630\leq a \leq 1135$ km, taken every $5$ km. The other elements were taken as before for region $1$. 
This combination of initial conditions resulted in a total of approximately $112,000$ particles distributed within region $2$.

\begin{figure*}
\mbox{%
\subfigure[]{\includegraphics[height=2.9cm]{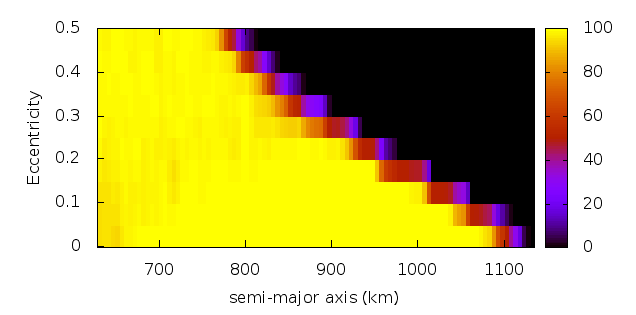}}\qquad
\subfigure[]{\includegraphics[height=2.9cm]{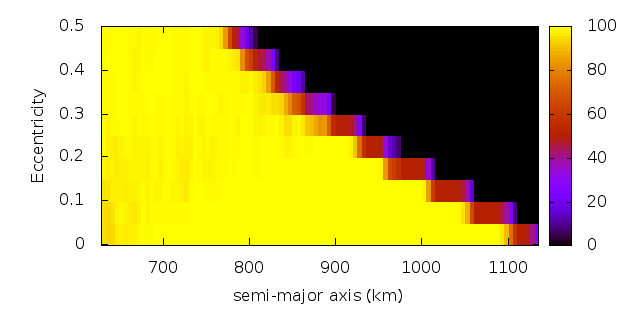}}
\subfigure[]{\includegraphics[height=2.9cm]{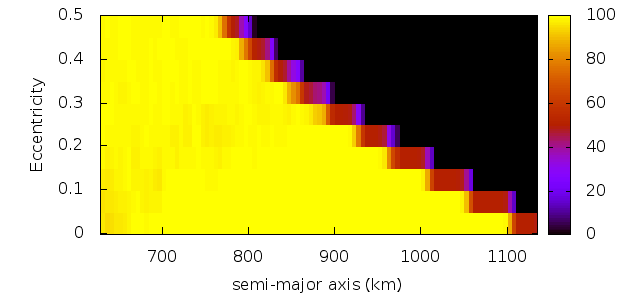}}}
\mbox{%
\subfigure[]{\includegraphics[height=2.9cm]{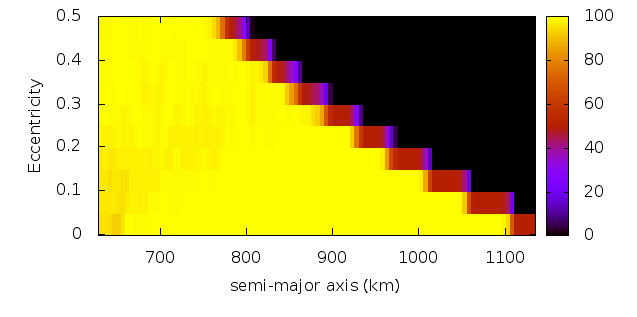}}\qquad
\subfigure[]{\includegraphics[height=2.9cm]{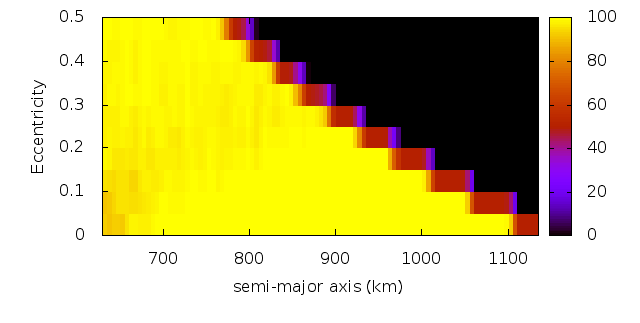}}
\subfigure[]{\includegraphics[height=2.9cm]{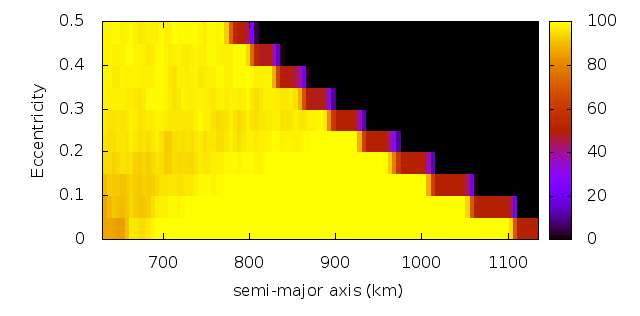}}}
\mbox{%
\subfigure[]{\includegraphics[height=2.9cm]{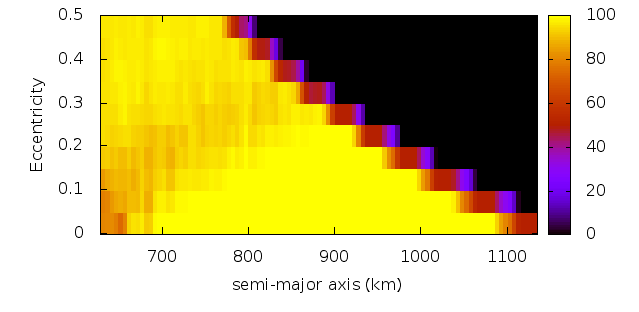}}}
\caption{Diagrams $(a \times e)$ of stability of  region $2$ for the triple system (45) Eugenia, for a time span of $2$ years. In this region the particles orbit Eugenia with orbital inclinations: a) $I=0^{\circ}$, b) $I=30^{\circ}$, c) $I=60^{\circ}$, d) $I=90^{\circ}$, e) $I=120^{\circ}$, f) $I=150^{\circ}$, g) $I=180^{\circ}$.
The color-coded scale indicates the percentage of particles that survives in this region, going from $0\%$ (instability) until $100\%$ (stability).}
\label{fig_histogram2}
\end{figure*}

The results for region $2$ are presented in the diagrams of Fig. \ref{fig_histogram2}. 
As found for region $1$, we see that approximately the whole region $2$ is stable. The instability appears only for the orbits that 
cross the orbit of Petit-Prince.
The results for this region also show that, once again, there are no effects from the inclination in the survival of the particles, 
as occurred in region 1. It is one more time different from what happened for the system 2001 SN263, as detailed in Fig. $7$ 
of \cite{araujo2012}, where the inclination has a strong effect in the evolution of the orbits of the particles, 
with a maximum at 75 degrees. The study of retrograde orbits made in \cite{araujo2015} 
showed a very large region of stable/retrograde and unstable/direct orbits (Fig. $8$). 
It means that the (45) Eugenia system does not have this option for selecting orbits for a space mission also in this region. 
The reason is the same explained before. The weak Kozai effect due to the smaller mass ratio and larger distances of the third perturbing body.

\subsection{Region 3}
\label{sec_r3}

Region $3$ was defined as the region around Princesse, going from $2.6$ km ($1.04$ radius of Princesse) until $15$ km (Hill's radius of Princesse with respect to Eugenia)

The particles were distributed within this region orbiting Princesse with $630\leq a \leq 1135$ km, taken every $5$ km. 
The other elements were taken as before. 
This combination of initial conditions resulted in a total of approximately $14,300$ particles distributed within region $3$.

\begin{figure*}
\mbox{%
\subfigure[]{\includegraphics[height=2.9cm]{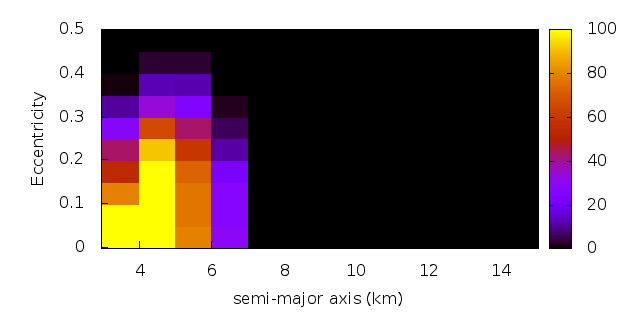}}\qquad
\subfigure[]{\includegraphics[height=2.9cm]{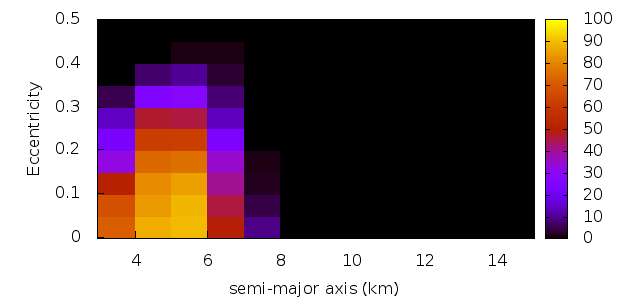}}
\subfigure[]{\includegraphics[height=2.9cm]{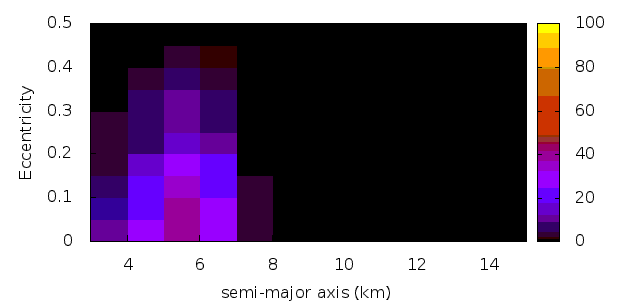}}}
\mbox{%
\subfigure[]{\includegraphics[height=2.9cm]{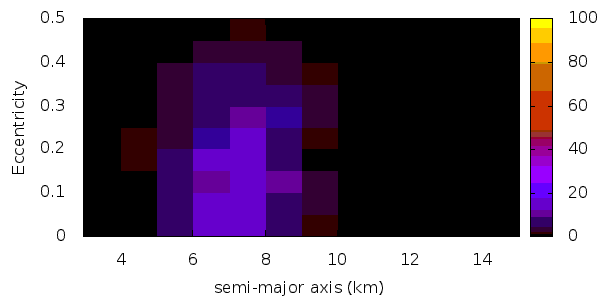}}\qquad
\subfigure[]{\includegraphics[height=2.9cm]{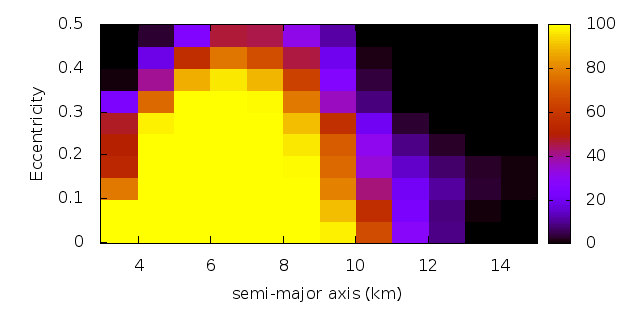}}
\subfigure[]{\includegraphics[height=2.9cm]{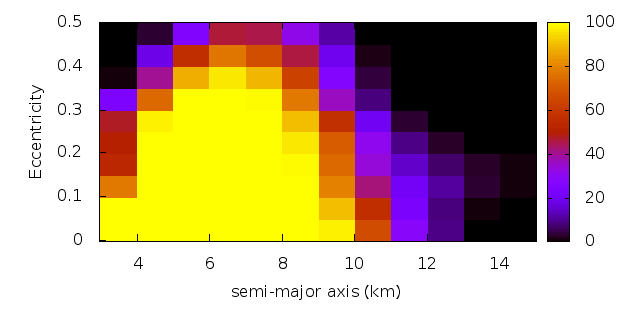}}}
\caption{Diagrams $(a \times e)$ of stability of  region $3$ for the triple system (45) Eugenia, for a time span of $2$ years. 
In this region the particles orbit Princesse with orbital inclinations: a) $I=0^{\circ}$, b) $I=30^{\circ}$, c) $I=60^{\circ}$, 
d) $I=120^{\circ}$, e) $I=150^{\circ}$, f) $I=180^{\circ}$.
The color-coded scale indicates the percentage of particles that survives in this region, going from $0\%$ (instability) until $100\%$ (stability).
The diagram for $I=90^{\circ}$ is not presented, since all the particles in this region with this inclination were lost in $2$ years (collisions or ejections).}
\label{fig_histogram3}
\end{figure*}

The results for region $3$ is presented in the diagrams of Fig. \ref{fig_histogram3}. 
The results for this region show that there are strong effects of the inclination in the survival of the particles. 
The Kozai effect is now present because the third body perturbation is now very strong since the mass ratio between 
the main body and the perturbing body is also $4.46 \times 10^{-5}$. Therefore, orbits with higher inclination are more stable than 
the planar ones, when out of the region where there is the kozai resonance action. 
Even though, the results are still different from the ones obtained for the system 2001 SN263, shown in \cite{araujo2012}. 
For the system 2001 SN263 no stable orbits were found around the closer smaller body, due to the very strong perturbation coming 
from the central body. Even retrograde stable orbits do not exist, as shown in \cite{araujo2015}. 
It means that the (45) Eugenia system has excellent natural orbits for a space mission to observe the smaller body closer to the central body, 
opposite to the 2001 SN263 system, which requires orbital control to observe this body. 

It is clear, from Fig. \ref{fig_histogram3}, that retrograde circular orbits with semi-major radius from $7$ km to $10$ km, 
depending on the initial eccentricity, are excellent 
for an exploration mission to this system, because they are stable while the equivalent direct orbits (same size and shape) are unstable

\subsection{Region 4}
\label{sec_r4}

Region $4$ was defined as the region around Petit-Prince, going from $3.6$ km ($1.03$ radius of Petit-Prince) until 
$28$ km (Hill's radius of Petit-Prince with respect to Eugenia)

The particles were distributed within this region orbiting Petit-Prince with $3.6 \leq a \leq 28$ km, taken every $5$ km. 
The other elements were taken as before. 
This combination of initial conditions resulted in a total of approximately $27,500$ particles distributed within region $4$.

The results for this region show smaller effects from the inclination in the survival of the particles, when compared to region $3$. 
The Kozai effect is present, but it is very small. Although the third body still has a larger mass, the distance between the bodies are much larger, 
compared to region $3$. 
The results for the case of the system 2001 SN263, shown in Fig. $8$ of the 
paper by \cite{araujo2012}, indicate much larger effects in the most distant body. 
It happens due to the much smaller distances involved in the system 2001 SN263, which contributes to higher third-body perturbations. 
The study of the retrograde orbits showed larger regions of stable orbits, as occurred for the system 2001 SN263, as shown in \cite{araujo2015}, Fig. $8$. 
It means that the (45) Eugenia system has excellent locations to place a spacecraft willing to observe the smaller body which is far away 
from to the central body. 

\begin{figure*}
\mbox{%
\subfigure[]{\includegraphics[height=2.9cm]{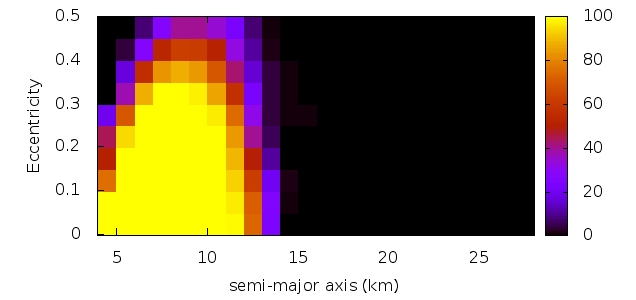}}\qquad
\subfigure[]{\includegraphics[height=2.9cm]{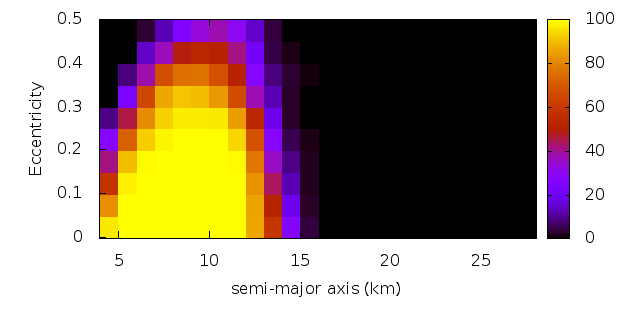}}
\subfigure[]{\includegraphics[height=2.9cm]{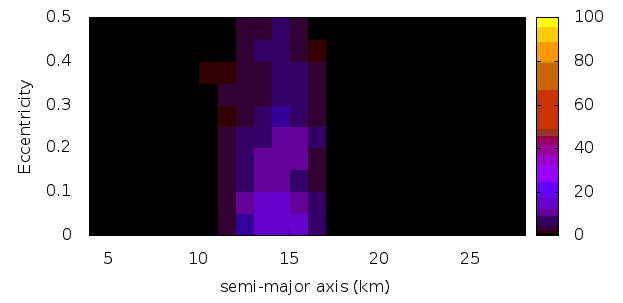}}}
\mbox{%
\subfigure[]{\includegraphics[height=2.9cm]{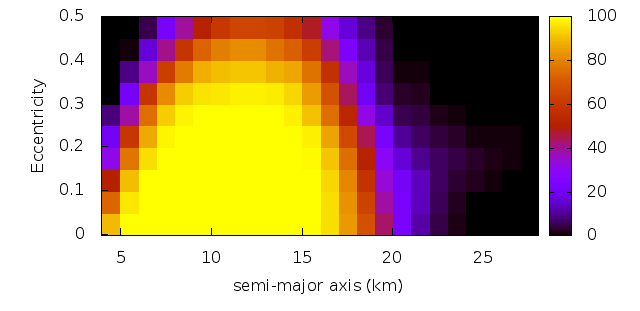}}\qquad
\subfigure[]{\includegraphics[height=2.9cm]{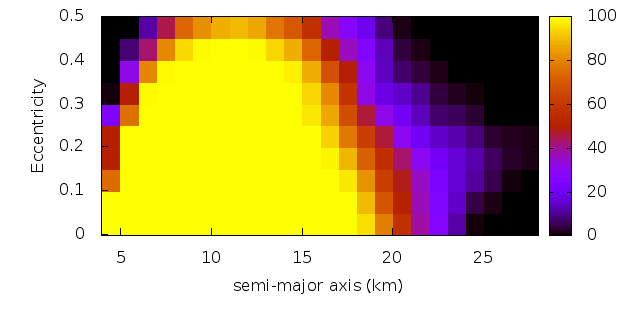}}}
\caption{Diagrams $(a \times e)$ of stability of  region $4$ for the triple system (45) Eugenia, for a time span of $2$ years. 
In this region the particles orbit Petit-Prince with orbital inclinations: a) $I=0^{\circ}$, b) $I=30^{\circ}$, c) $I=120^{\circ}$, 
d) $I=150^{\circ}$, e) $I=180^{\circ}$.
The color-coded scale indicates the percentage of particles that survives in this region, going from $0\%$ (instability) until $100\%$ (stability).
The diagrams for $I=60^{\circ}$ $I=90^{\circ}$ are not presented, since all the particles in this region with this inclination were lost in $2$ years (collisions or ejections).}
\label{fig_histogram}
\end{figure*}

The results for regions $1$, $2$, $3$ and $4$ showed the existence of stable orbits in the regions around the three bodies of the system. Those results are different from what 
was obtained when studying the 2001 SN263 system, where no stable orbits were found near the smallest body of the system. 
Several other important differences were found. In particular, there are no places around the main body where there are stable retrograde 
orbits and that are unstable in the prograde orbits. It means that this strategy to choose orbits is 
not available to observe the main body. On the other hand, situations like that are found around the smaller bodies of the system.
In general, the lower perturbation level that occur in all 
the regions of the (45) Eugenia system makes this system much easier to place a spacecraft, from the point of view of longer 
duration missions requiring less orbital control. This is a key point in deep space missions, due to the technical 
difficulties of carrying more fuel and making frequent orbital maneuvers at this distance.

It is clear, from Fig. \ref{fig_histogram}, that retrograde circular orbits with semi-major radius from $14$ km to $18$ km, 
depending on the initial eccentricity, are excellent 
for an exploration mission to this system, because they are stable and the direct orbits of the same size are unstable.

\section{Conclusions}

A detailed study of the (45) Eugenia system was made, with the goal of finding stable orbits in the several regions of this triple asteroid system. 
Those stable orbits are important to indicate regions of possible accumulation of dust, or even good locations to looking for a new member of 
the family of the asteroids system. In that sense, stable regions were found, as a function of the semi-major axis, eccentricity and inclination of the orbits, 
around all the bodies that composes the system. 

In general, the results showed that the (45) Eugenia system is much less perturbed than the triple system 2001 SN263, which was studied before. 
The integral of the perturbing forces quantified this fact in numbers. In particular, the Kozai effect of destroying inclined orbits was not 
present in orbits around the central body, due to the small mass ratio and large distance of the third-body perturbation. 
However, this effect is present in orbits around the two satellites.

Those stable orbits are also important in terms of astronautical applications, because regions of the space that has unstable/direct and 
stable/retrograde orbits are good locations to place a spacecraft, since they give stability for the orbit and a low risk of collisions with particles.
A new type of integral index was defined, taking into account the perturbations coming from the other bodies of the systems compared to the body that the 
spacecraft is orbiting. It showed a comparison of the perturbation levels, indicating that orbits around the main body of the (45) Eugenia system are 
about $10^4$ times less perturbed than the equivalent orbits on the 2001 SN263 system. Regarding the companion bodies, 
the internal body of the (45) Eugenia system is also about $10^4$ times less perturbed than the equivalent body of the 2001 SN263 system. 
For the external bodies, the levels of perturbations are similar for both systems.

The results showed that orbits to locate a spacecraft using the difference in the stability 
for direct and retrograde orbits do not exist in the (45) Eugenia system for the regions near the central body and between the orbits of 
the two smaller bodies, as happened for the 2001 SN263 system. But, on the opposite side, such regions were found around the two smaller 
bodies of the system, covering regions where no stable orbits were found for the 2001 SN263 system. It means that, in terms of observing the 
smaller bodies of the system, the triple asteroid (45) Eugenia offers better locations to place a spacecraft, with cleaned stable retrograde orbits.
The numerical results showed that orbits with semi-major radius from $7$ km to $10$ km are very good choices for orbits around Princesse and 
orbits with semi-major radius from $14$ km to $18$ km are very good choices for orbits around Petit-Prince.

\section{Acknowledgments}
This work was funded by CNPq Procs. 150378/2015-7, Proc. 301338/2016-7 , and by FAPESP Procs. 2016/14665-2 and 2011/08171-3. 
This support is gratefully acknowledged.

\renewcommand{\refname}{REFERENCES}

\label{lastpage}


\begin{thebibliography}{2}
\bibitem[\protect\citeauthoryear{}{}]{b}


\bibitem[\protect\citeauthoryear{Araujo et al.}{2012}]{araujo2012}
Araujo, R. A. N.; Winter, O. C.; Prado, A. F. B. A.; Sukhanov, A., 2012, MNRAS, 423, 4, 3058.

\bibitem[\protect\citeauthoryear{Araujo et al.}{2015}]{araujo2015}
Araujo, R. A. N.; Winter, O. C.; Prado, A. F. B. A., 2015, MNRAS, 449, 4, 4404.

\bibitem[\protect\citeauthoryear{Beauvalet \& Marchis}{2014}]{Beauvalet2014}
Beauvalet, L.; Marchis, F., 2014, Icar, 241, 13.

\bibitem[\protect\citeauthoryear{Carvalho, Moraes \& Prado}{2014}]{carvalho2014}
Carvalho, J. P. S. ;  Moraes, R. V.; Prado, A. F. B. A., 2014, MPE, 2014, 529716, 10 pp.
 
 \bibitem[\protect\citeauthoryear{Chanut, Winter \& Tsuchida}{2014}]{chanut}
 Chanut, T.G.G.; Winter, O.C.; Tsuchida, M., 2014, MNRAS, 438, 2672.
 
 
\bibitem[\protect\citeauthoryear{Everhart}{1985}]{everhart}Everhart, E. 
An efficient integrator that uses Gauss-Radau spacings. In Dynamics of comets: Their origin and evolution, 
Eds. A. Carusi Carusi and G. B. Valsecchi, D.Reidel Publishing Company (Holanda), p. 185-202, 1985.

\bibitem[\protect\citeauthoryear{Fujita et al.}{2011}]{fujita}
Fujita, K.; Yamamoto, M.; Abe, S.; Ishihara, Y.; Iiyama, O.; Kakinami, Y.; Hiramatsu, Y.; Furumoto, M.; Takayanagi, H.; Suzuki, T.; 
Yanagisawa, T.; Kurosaki, H.; Shoemaker, M.; Ueda, M.; Shiba, Y.; Suzuki, M.
An Overview of JAXA's Ground-Observation Activities for HAYABUSA Reentry. 
Publications of the Astronomical Society of Japan, Vol.63, No.5, pp.961-969, 2011.

\bibitem[\protect\citeauthoryear{Hergenrother et al.}{2014}]{b15}
Hergenrother, Carl W.; Barucci, M. A.; Barnouin, O.; Bierhaus, B., et al., 2014, eprint arXiv:1409.4704, 116 pages.

\bibitem[\protect\citeauthoryear{Horizons-JPL}{}]{horizons} HORIZONS Web-Interface. Available in
\url{http://ssd.jpl.nasa.gov/horizons.cgi#top}.
Accessed in: 15th February, 2017.

 \bibitem[\protect\citeauthoryear{Jiang et al.}{2016}]{jiang}
 Jiang,Y.; Zhang, Y.; Baoyin, H.; Li, J., 2016, ApSS, 361,9,306.
 
 \bibitem[\protect\citeauthoryear{Lara}{2016}]{lara2016}
Lara, M., 2016, J. Guid., Cont., and Dynam, 39 (9), 2157.

\bibitem[\protect\citeauthoryear{Marchis et. al.}{2010}]{marchis}
Marchis, F.; Lainey, V.; Descamps, P.; Berthier, J.; et al., 2010, Icarus, 210, 2, 635.

\bibitem[\protect\citeauthoryear{Murray and Dermott}{1999}]{murray} Murray, D.C.; Dermott, S.F. Solar System Dynamics. Cambridge University Press. 1999. 

\bibitem[\protect\citeauthoryear{Oliveira \& Prado}{2014}]{olivprado2014}
Oliveira, T.C.; Prado, A.F.B.A., 2014, Acta Astronau. 104, 350.

\bibitem[\protect\citeauthoryear{Oliveira, Prado \& Misra}{2014}]{oliveiraetal2014}
Oliveira, T.C.; Prado, A.F.B.A.; Misra, A.K., 2014, Adv. in the Astronau. Sci. 152, 3081.

\bibitem[\protect\citeauthoryear{Prado}{2014}]{prado2014} Prado, A. F. B. A., 2014,
Advances in Space Research, 53, 5, 877.

\bibitem[\protect\citeauthoryear{Prado}{2013}]{prado2013}  
Prado, A. F. B. A., 2013, MPE, 2013, ID 415015, pp. 1.

\bibitem[\protect\citeauthoryear{Sanchez, Prado \& Yokoyama}{2014}]{sanchesetal2014}
Sanchez, D. M.; Prado, A.F.B.A.; Yokoyama, T., 2014, Adv. in Space Res. 54, 1008.

\bibitem[\protect\citeauthoryear{Sanchez, Howell \& Prado}{2016}]{sanchesetal2016}
Sanchez, D.M.; Howell, K.C.; Prado, A.F.B.A., 2016, AAS/AIAA Spaceflight Mech. Meeting, NAPA, CA, February 14-18, Feb (2016).

\bibitem[\protect\citeauthoryear{Santos et al.}{2015}]{santosetal2015}
Santos, J.C.; Carvalho, J.P.S.; Prado, A.F.B.A.; Moraes, R.V., 2015, J. Physics Conf. Series.  641, 012011 Oct. (2015)

\bibitem[\protect\citeauthoryear{Short et al.}{2016}]{shortetal2016}
Short, C.; Howell, K.; Haapala, A.; Dichmann, D., 2016, J. Astronau. Sci., 64, 2, 156.

\bibitem[\protect\citeauthoryear{Sukhanov et al.}{2010}]{sukhanov}S
ukhanov, A.A., Velho, H.F.C., Macau, E.E., Winter, O.C., 2010, Cosmic Research, 48, 5, 443.

\bibitem[\protect\citeauthoryear{Yoshikawa}{2006}]{b10} Yoshikawa,  M. et al.,
Technologies for future asteroid exploration: What we learned from hayabusa mission. Spacecraft Reconnaissance of Asteroid and Comet Interiors, 
n.3038,2006.


\end{thebibliography}
\end{document}